\documentclass[twocolumn,aps,prb,longbibliography]{revtex4-2}

\usepackage{graphicx}
\usepackage{dcolumn}
\usepackage{bm}
\usepackage{amsmath}
\usepackage[utf8]{inputenc}
\usepackage[T1]{fontenc}
\usepackage{mathptmx}
\usepackage{etoolbox}
\usepackage{color}
\usepackage{ulem}

\begin{document}

\title{\textcolor{blue}{Quantitative analysis of vectorial torques in thin 3\textit{d} Co ferromagnet using orbital-spin conversion.}}

\author{B. Bony}
\author{S. Krishnia}
\author{Y. Xu}
\author{S. Collin}
\author{A. Fert}
\author{J.-M. George}
\affiliation{Laboratoire Albert Fert, CNRS, Thales, Universit\'e Paris-Saclay, 91767 Palaiseau, France}
\author{M. Viret}
\affiliation{Service de Physique de l'Etat Condens\'e, CEA, CNRS, Universit\'e Paris-Saclay, Gif-sur-Yvette, 91191, France}
\author{V. Cros}
\author{H. Jaffrès}
\affiliation{Laboratoire Albert Fert, CNRS, Thales, Universit\'e Paris-Saclay, 91767 Palaiseau, France}
 
\email{benjamin.bony@cnrs-thales.fr,henri.jaffres@cnrs-thales.fr}

\date{\today}

\begin{abstract}
        Recent findings in orbitronics pointed out large current-induced torques originating, in the current understanding, from incident orbital currents. These are generated by orbital Rashba-Edelstein effect (OREE) produced at the interface between some light metal and oxides films \textit{e.g.} by naturally oxidized copper layer (Cu*).  In the present work, by using second harmonic Hall techniques, we determine the ratio of orbital \textit{vs} spin currents exerting torques on thin transition metals Co ferromagnet in systems using an orbit-to-spin Pt converter as interlayer with Cu*. Our results quantifying damping like torques show that both orbital and spin currents are enhanced in these systems. Moreover, the experimental determination of the decoherence length in a sample series with varying Co thickness clearly demonstrates the interfacial generation of the orbital currents in Cu* by Orbital Rashba-Edelstein effects (REE) leading to subsequent magnetic torque in Co over a typical lengthscale of several nanometers. 
\end{abstract}


\maketitle 

\section{Introduction}

During the last decades, the fast development of spintronics has led to the discovery of rich concepts involving spin-transfer (STT) and spin-orbit torques (SOT). These aim to enhance spintronic device performances with the use of pure spin currents \textit{via} spin-orbit interaction. A clear benefice is an electrical control and manipulation of the magnetization, \textit{e.g.} in SOT-MRAM technology~\cite{manchon2019,miron2010,miron2011} constituting nowadays an energy-efficient alternative to semiconductor based memories. For these perspectives, a better understanding of inherited fundamental concepts as charge-spin conversion phenomena such as spin Hall (SHE) or Rashba-Edelstein (REE) effects and reciprocals is mandatory. 

Very recently, the physics of orbital angular momentum (OAM) has been the focus of important breakthroughs in this topics. The framework of orbital torque (OT)~\cite{Lee2021,Go2018, go2020,GoEPL} relies on the generation of polarized OAM, possibly its transport over nanometric distance, and its conversion into spin angular momentum through spin-orbit (SO) interaction. Indeed, several mechanisms may be at the origin of the OAM production such as the  orbital hall effect (OHE)~\cite{Go2018, go2020,salemi2022,Bauer2024,Choi2023}, the orbital Rashba-Edelstein effect (OREE)~\cite{Go2017, johansson2021,ding2022,johansson2023,ElHamdi2023} and possibly non-local or current gradient effects at the vicinity of specific interfaces~\cite{valet2024}. Those have been shown to overcome their analog pure spin effects in some cases~\cite{Leandro2022, GiganticJo2018, Sala2022GiantOHE}. In order to exert a magnetic torque on a ferromagnet \textit{via} exchange processes, the orbital currents have to be converted into spin using spin-orbit interactions. The orbit-to-spin conversion may occur in the ferromagnetic (FM) itself during the torque process or into an intermediate non magnetic layer (NM). In the case of light materials, OT offers comparable efficiencies to Pt~\cite{Otani2023, Sala2022GiantOHE} a reference SHE material. By combining light metals with heavy metal orbit-to-spin converters, a large enhancement of torques has been thus demonstrated~\cite{lee2021efficient, krishnia2024quantifying, Ding2020, Santos2023, gupta2024harnessing}. Moreover, recent investigations showed the occurrence of a strong OREE~\cite{nikolaev2024} at the interface between Al and an ultrathin Co producing a strong increase of the field-like torque by one order of magnitude~\cite{krishnia2023nano}.

Recent works focused more particularly on the role of  O \textit{p}-type and Cu \textit{d} orbital hybridization at the interface between metallic Cu and CuO to produce OREE~\cite{go2021CuCuO} and subsequent OAM polarization. Numerous experiments showed that the natural oxidation of Cu (noted Cu*) led to large torque enhancement in bilayers~\cite{an2016spin, Otani2021, Otani2023, ding2022, huang2023reversalLongrange, Ding2024GdCo, ding2024mitigation}, in lateral geometry devices~\cite{otani2025} or with the use of Pt spacer for its role in the orbit-to-spin conversion~\cite{Ding2020, Santos2023, krishnia2024quantifying, CuOMagnons, sahoo2024efficient}. Nevertheless, the actual composition of Cu* remains complex with the presence of three possible oxidation degrees for Cu (Cu$^0$, Cu$_{2}$O and CuO)~\cite{an2016,krishnia2024quantifying} and a strong dependence on the Cu crystallinity\cite{Otani2023}. The dependence of the OT on the Cu thickness may indicate the pure interfacial origin or not, possibly confirming the occurrence of OREE at the interface between the light metal and the oxide film~\cite{Otani2023}.  

One one hand, questions arise about the typical coherence length to expect for a non-equilibrium orbital current in paramagnets over distance larger than typically 1~nm~\cite{kelly2023}. However, long range magnetic torques have thus been demonstrated in non-magnetic light elements (Ti, Cr) via OHE~\cite{Hayashi2023, Choi2023, lee2021efficient,Chiba2024} or more recently with non-local measurments in Cu|CuO$_x$ lateral devices~\cite{otani2025}. On the other hand, orbital currents in ferromagnets have been predicted to possibly experience longer propagation length than spin currents due to the specific orbital texture in the band structure and the presence of hot-spots transport in the reciprocal k-space ~\cite{go2022longrangeorbitalmagnetoelectrictorque}. A clear understanding of how far the orbital polarization can be active on a magnetization is still under debate~\cite{Lee2024}. The dephasing (or decoherence) of the spins penetrating a magnetic materials around its exchange fields is known to be really short ($\sim 1$~nm or less) in 3\textit{d} transition metals~\cite{Stiles2002,taniguchi2008penetration, ghosh2012penetration, krishnia2023nano}. As the OAM interact with the exchange field only through spin-orbit coupling, its propagation in magnetic elements may remain longer than for the spins reaching about 20~nm in Ni~\cite{Hayashi2023}, 5~nm in Co~\cite{UMRGambardella2022} or in CoFe~\cite{ding2024mitigation} and CoFeB~\cite{huang2023reversalLongrange}. A precise evaluation of the characteristic lengthscale for the torques may differentiate the respective orbital and spin two contributions. Such methodology has been applied to Co|Pt systems~\cite{Sala2022GiantOHE} for which the increase of the torque above the characteristic spin decoherence length has been assigned to the OHE of Pt~\cite{Sala2022GiantOHE,kelly2024}, the positive sign of the SOC in Pt leading to additive orbital and spin actions. Whereas this method has been successfully applied to NM|FM bilayers involving either light or heavy materials, it has never been applied in systems integrating an intercalated Pt spacer. This appears as a direct way to determine if the associated torque enhancement truly originates from orbital-spin conversion, and to possibly give new insights on the interplay between spin and orbital degrees of freedom in SOT devices. 

In our previous study performed on Co|Pt|Cu*, we pointed out that the maximum enhancement arises for 4~nm Pt~\cite{krishnia2024quantifying}. In this work, we aim at \textit{i}) evaluating the magnetic torque and corresponding decoherence length through Co thickness dependence in Co|Pt|Cu* series, \textit{ii})
determining experimentally the ratio of orbital and spin contributions in the torque enhancement due to the intercalated Pt spacer, \textit{iii}) checking the interfacial origin of the orbital current in Cu*. In this article, we present our results reporting on the extraction of the decoherence length in Co by performing second harmonic hall voltage measurements on different series: 
Co(t$_{Co}$)|Pt(3), 
Co(t$_{Co}$)|Pt(3)|Cu(3)*, Co(t)|Pt(4)|Cu(3)* and in series with varying Cu thickness: Co(2)|Pt(3)|Cu(t$_{Cu}$)*, Co(2)|Cu(t$_{Cu}$)*.

\section{\label{sec:level1}Material systems.}

\subsection{Sample growth and magnetic characterizations}

The investigated sample series were deposited by dc magnetron sputtering on thermally oxidized Si|SiO$_2$(280~nm) substrates. The choice of depositing Co onto SiO$_2$, that is without the use of Ta|Pt bottom buffer (or any other buffer layers), is to avoid any generation of spurious spin-Hall currents injected at the opposite Co interface. Such growth procedure leads generally to a small increase of the layer resistivity due to slightly larger roughness. However a set of Co|Pt optimized reference series grown on Ta(5)|Pt(8) buffer layers can be found in former work\cite{krishnia2023nano} and compared with the equivalent series grown onto SiO$_2$ to check our approach and in particular the validity of our data analyses. The samples are polycrystalline and magnetized in the plane. The oxidation of Cu has been characterized in our previous work~\cite{krishnia2024quantifying} by X-ray Photoelectron Spectroscopy (XPS) experiments and STEM cross-section analyses showing the presence of a separate metallic copper Cu(0) and oxidized copper with different degree of oxidation (CuI, CuII) in the naturally oxidized 3~nm Cu layer. The resistivities and THz-TDS (time domain spectroscopy) data are compatible with an equivalent 1~nm metallic Cu and 2~nm of oxidized Cu for a 3~nm pristine Cu overlayer. Details of the growth and characterizations of Cu* are given in the Ref.~\cite{krishnia2024quantifying}.

\begin{figure*}
\includegraphics[width=0.75\textwidth]{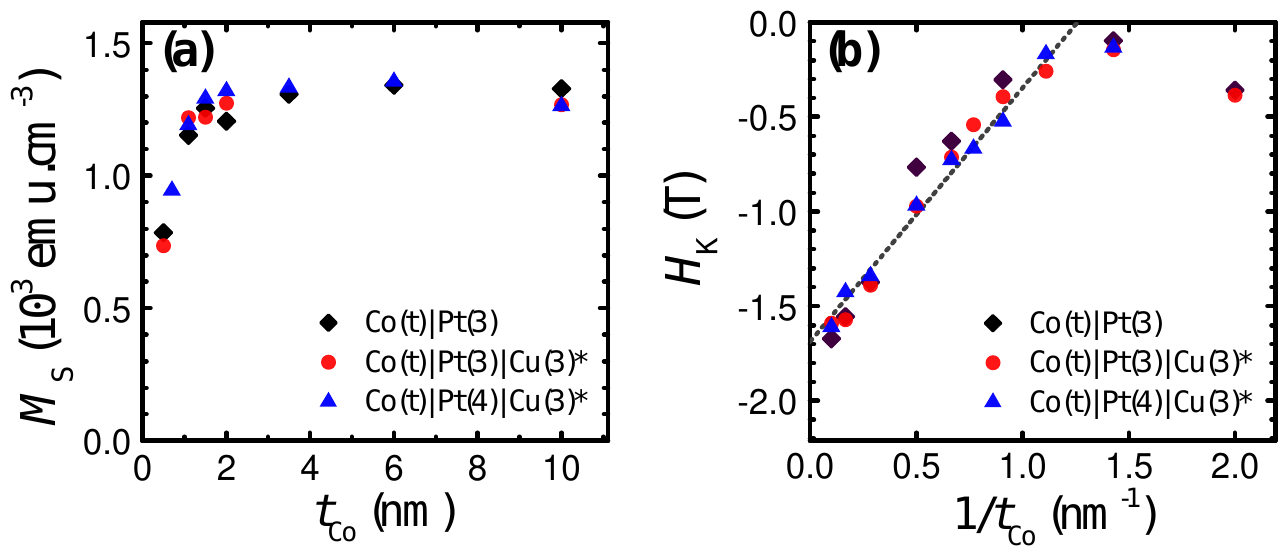}
\caption{\label{fig:Ms}Cobalt thickness dependence for Co($t_{Co}$)|Pt(3), Co($t_{Co}$)|Pt(3)|Cu(3)* and Co($t_{Co}$)|Pt(4)|Cu(3)* of a) magnetization saturation $M_S$ measured by SQUID, b) anisotropy fields $H_K$ measured from first harmonic of the Hall voltage by out-of-plane field sweeping. The measurements were performed at 300~K.}
\end{figure*}

\vspace{0.1in}

The saturation magnetization values ($M_s$) for the different series has been obtained by SQUID measurements at room temperature for both sample series (Fig.~\ref{fig:Ms}a). We find quantitatively a similar behavior for the different sample series. For Co layer thicknesses $t_{Co} > 1.4$~nm, $M_s\simeq 1300$~emu/cm$^3$ very close to the expected value for bulk Co at room temperature. For thinner Co, (typically for $t_{Co} < 1.4$~nm) $M_s$ gradually decreases down to $M_s\simeq 750$~emu/cm$^3$ for $t_{Co} < 0.5$~nm. Such reduction of $M_s$ for $t_{Co}$ below the nanometer scale have already been observed in Co|Al systems, possibly assigned to local strain effects~\cite{krishnia2025}. However, this may also indicate superparamagnetic properties of ultrathin Co layers. The effective in-plane anisotropy field $H_K^{eff}=4\pi M_s-2K_s/\left(M_s t_{Co}\right)$ with $K_s$ the surface anisotropy originating from the SiO$_2$|Co|Pt interfaces is plotted on Fig.~\ref{fig:Ms}b \textit{vs.} the inverse Co thickness 1/t$_{Co}$. First, the negative sign of the anisotropy for all the samples indicates an in-plane anisotropy. Second, the intercept of $H_K^{eff}$ with the ordinate axis corresponding to bulk Co (infinite thickness) is about $1.7$~T as expected ($4\pi M_s\simeq 1.7$~T). Secondly, the well-defined slope of $H_K^{eff}$ \textit{vs.} 1/t$_{Co}$ for 1/t$_{Co}$ < 1~nm$^{-1}$ (t$_{Co}$ > 1~nm) indicates the pretty good quality of Co and its interface for this thickness range in agreement with the information given by the $M_s$. Thirdly, the slope equivalent for the three sample series indicates a surface anisotropy strength $K_s\simeq 0.98$~erg/cm$^2$ (0.98 $\simeq$ mJ/m$^2$) in agreement with Ref.~\cite{guo2006survey}.

\subsection{\label{sec:level2electrical}Spin currents characterizations by Anomalous Hall effects (AHE).}

Anomalous Hall effects (AHE) can be used to probe the spin-dependent scattering and spin-current properties at the vicinity of the interfaces~\cite{Dang2020,krishnia2023nano,krishnia2024quantifying}. To realize these characterizations, we have fabricated $W=5~\mu m$ wide and $L=50~\mu m$ long Hall crosses by optical lithography followed by an ion beam (IBE) etching step. The same Hall bars are used for torque extraction. The dependence of the longitudinal resistance ($R_{xx}$) \textit{vs.} the Co layer thickness ($t_{Co}$) including the Pt spacer (of thickness $t_{Pt}$) gives an estimate of the material resistivity. It approaches $\rho_{Pt}\simeq 50~\mu\Omega$.cm for Pt in a large thickness scale and $\rho_{Co}$ in the range 40-100~$\mu\Omega$.cm for Co depending on its thickness. 

\begin{figure*}
\includegraphics[width=0.95\textwidth]{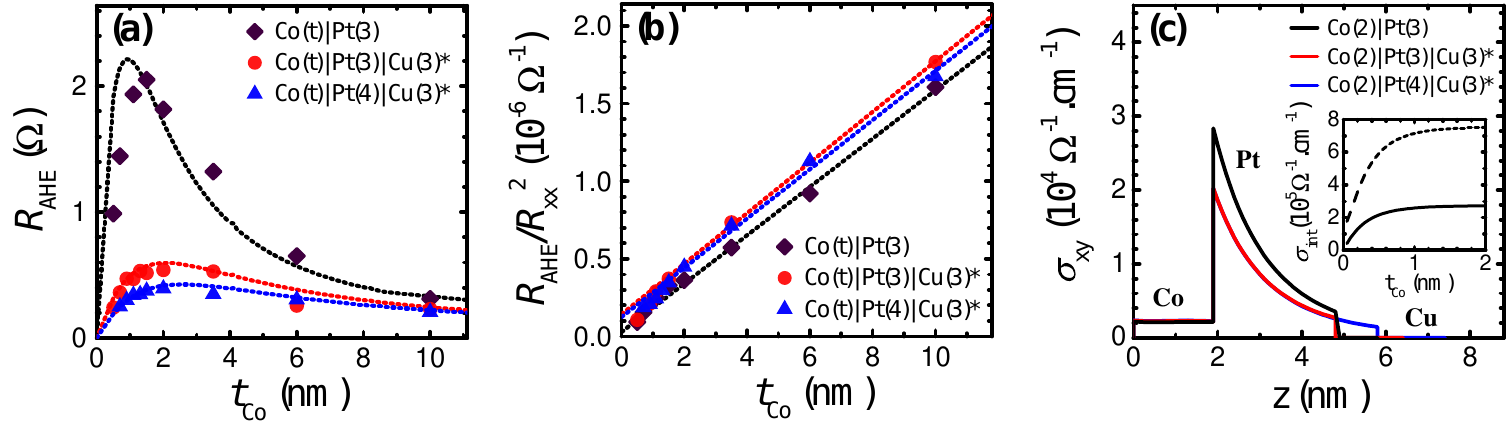}
\caption{\label{fig:AHE}Cobalt thickness dependence for Co($t_{Co}$)|Pt(3), Co($t_{Co}$)|Pt(3)|Cu(3)* and Co($t_{Co}$)|Pt(4)|Cu(3)* series of a) the anomalous Hall resistance (AHE) $R_{AHE}$ measured from first harmonic of the Hall voltage by out-of-plane field sweeping, b) $R_{AHE}/R_{xx}^{2}$ ratios showing a linear behavior owing to the intrinsic spin-Hall conductivity of Co. Measurements were performed at 300~K. c) Calculated layer-resolved local transverse conductivity $\sigma_{xy}(z)$ for the three sample series showing enhancement of the AHE in the Pt layers at the scale of the mean-free path (1.5~nm). Inset: Calculated current spin-polarization in Co ($\mathcal{P}_s^{Co}$, dashed line) and in Pt ($\mathcal{P}_s^{Pt}$, straight line) \textit{vs.} the Co thickness ($t_{Co}$).}
\end{figure*}

The parallel resistor model for the spin-dependent conduction includes possible current shunting in Pt and Cu* layers as far as Cu* is not fully oxidized. This can be quantified from the reduction of the anomalous Hall effect (AHE) compared to the reference ('ref') samples free of such current shunting according to $R_{AHE}\left(\zeta\right)=\zeta^2 ~R_{AHE}^{ref}$  ($\zeta=\frac{R_{xx}}{R_{xx}^{ref}}<1$ is less than unity). In Fig.~\ref{fig:AHE}a, we display the AHE values measured on the following sample series: Co($t_{Co}$)|Pt(3) (ref), Co($t_{Co}$)|Pt(3)|Cu(3)* and Co($t_{Co}$)|Pt(4)|Cu(3)*. The continuous curves are the result of our fitting procedure following our theory established for multilayers~\cite{Dang2020}. The increase of AHE observed for Co thinner than 2~nm describes the gradual rise of the spin-current polarization with $t_{Co}$ from both the bulk (parameter $\beta\simeq 0.6\pm 0.05$) and the spin-dependent interfacial scattering (interfacial spin-scattering asymmetry $\gamma_{Co/Pt}=0.4$ from Ref.~\cite{Dang2020}). The decrease of AHE at large $t_{Co}$ observed for the three series follows the characteristic 1/t$_{Co}$ variation of the Co film resistance. Comparing the three different series, the smaller AHE magnitude for Co($t_{Co}$)|Pt(3-4)|Cu(3)* compared to Co($t_{Co}$)|Pt(3) can be quantitatively understood by \textit{i)} the current shunt in Cu(3)* with a native 1~nm thick metallic Cu layer whereas \textit{ii)} the difference in the AHE between Co($t_{Co}$)|Pt(3)|Cu(3)* and Co($t_{Co}$)|Pt(4)|Cu(3)* (red and blue curve) is explained by the additional current shunt introduced by Pt(4) compared to Pt(3), and \textit{iii)} an effective smaller Pt thickness (about 1.5~nm) with larger resistivity (50~$\mu\Omega$.cm) possibly due to Pt over oxidation for the Co(t$_{Co}$)|Pt3 series~\cite{krishnia2023nano}. To go further in the fitting analyses, we show in Fig.~\ref{fig:AHE}b the variation of $\frac{R_{AHE}}{R_{xx}^2}$ \textit{vs.} t$_{Co}$. For all the series, a linear variation of $\frac{R_{AHE}}{R_{xx}^2}$ is found according to:

\begin{eqnarray}
\left(\frac{W }{L}\right)^2\frac{R_{AHE}}{R_{xx}^2}\simeq\sigma_{AHE}^{Co}(t_{Co})~t_{Co}+\overline{T}_{Co/Pt}\theta_{SHE}^{Pt}\mathcal{P}^{Pt}_{s}(t_{Co})\sigma_{xx}^{Co}~\lambda_{Pt}
\end{eqnarray}
with $\mathcal{P}_{s}^{Pt}(t_{Co})$ the polarization of the spin current injected from Co to Pt, $\overline{T}_{Co/Pt}=\frac{0.4}{1-\gamma_{Co/Pt}^2}$ the average electronic transmission for majority spins at the Co/Pt interface ($\gamma_{Co/Pt}$ is the interfacial spin-asymmetry coefficient). The conductivity of Co, Pt and Cu was set to $\sigma_{xx}^{Co}=2.5\times 10^6~\Omega^{-1}$m$^{-1}$, $\sigma_{xx}^{Pt}=2\times 10^6~\Omega^{-1}$m$^{-1}$ and $\sigma_{xx}^{Cu}=5\times 10^6~\Omega^{-1}$m$^{-1}$ respectively. We set $\lambda_{Co}^{\uparrow}\simeq$ 5.7~nm and  $\lambda_{Co}^{\downarrow}\simeq$1.2~nm and $\lambda_{Pt}\simeq 1.5$~nm, the electron mean free path in Co (for majority $\uparrow$ and minority $\downarrow$ spins) and Pt respectively. The spin-memory loss parameter $\delta_{Co/Pt}$ is set to 0.4$\pm$0.05 corresponding to an effective spin loss of 0.5$\pm 0.1$. Such a typical linear variation observed at large  t$_{Co}$ should be associated to a spin-Hall conductivity for Co of $\sigma_{xy}^{Co}\simeq 180~\Omega^{-1}$cm$^{-1}$ corresponding to a mean Hall angle $\theta_{Co}^{AHE}=\frac{\sigma_{xy}^{Co}}{\sigma_{xx}^{Co}}\simeq 0.007$. The inset of Fig.~\ref{fig:AHE}c describes the gradual increase of the current spin-polarization $\mathcal{P}_s^{Co}$ \textit{vs.} $t_{Co}$ at the Co|Pt interface in the Co($t_{Co}$)|Pt|Cu trilayers inside Co (dashed line) and inside Pt (straight line). The typical lengthscale variation is the electronic mean free path for $\downarrow$ spin in Co. The difference between the two curves describes the spin-memory loss when electrons cross the interface. The linear variation observed on Fig.~\ref{fig:AHE}b is reached when $\mathcal{P}_s^{Co}$ reaches its saturation. Moreover, we have plotted in Fig.~\ref{fig:AHE}c the profile in depth of the Co(2)|Pt($t_{Pt}$)|Cu(3)* multilayers of the local transverse conductivity $\sigma_{xy}(z)$ arising from the proximity spin-current from Co. Here, one has considered the local spin-Hall angle in each separate layer, in particular in Pt showing the larger efficiency in the $\sigma_{xy}(z)$ values at the scale of its mean-free path (1.5~nm). The larger integrated value of $\sigma_{xy}^{Pt}(z)$ value in Pt for the Co(2)|Pt(3) bilayer is understood by a slightly larger resistivity of the top Pt in the bilayer compared to the trilayer. 

\section{\label{sec:level3} Quantification of Damping-like and Field-like torques and decoherence length}

Spin-orbit torques are evaluated by using standard second harmonic (f-2f) Hall measurements in the so-called in-plane ($\phi$) geometry corresponding to the rotation of the applied field in the plane (IP) of the stacks. We recently showed that out-of-plane (OOP) angular variation geometry give equivalent results~\cite{xu2024}. An oscillating current at frequency $f = 727$~Hz is injected while the transverse voltage $V_{xy}$ is measured. We first perform 1$^{st}$ harmonic measurement ($V^{1\omega}$) in out-of-plane field sweeping geometry to extract the AHE resistance (R$_{AHE}$) and magnetic anisotropy (H$_{K}$). We extract the planar Hall effect (PHE) voltage ($V^{1\omega}$) and resistance from the azimuthal field rotation. From the analysis of the 2$^{nd}$ harmonic signals ($V^{2\omega}$) we determine the respective damping-like (DL) and field-like (FL) effective fields. Note that we have checked on specific samples that the measurements and analyses of magnetic torques extracted from out-of-plane (OOP) angular geometry gave consistent results. According to Ref.~\cite{Hayashi2014Quantitative, Avci2014interplay}, the second harmonic voltage $V^{2\omega}$ writes: 

\begin{eqnarray}
R_{2\omega} = (R_{AHE}\times \frac{H_{DL}}{H_{K}+H_{ext}}+I\alpha\Delta T)\cos\phi +\nonumber \\ 
+ 2R_{PHE}\times(2\cos^{3}\phi-\cos\phi)\times\frac{H_{FL}+H_{Oe}}{H_{ext}}
\end{eqnarray}
with $H_{Oe}$ the Oersted field produced by the current and $\alpha$ the thermal coefficient giving rise to an additional voltage by Nernst effects. 

\begin{figure*}[ht]
\includegraphics[width=1\textwidth]{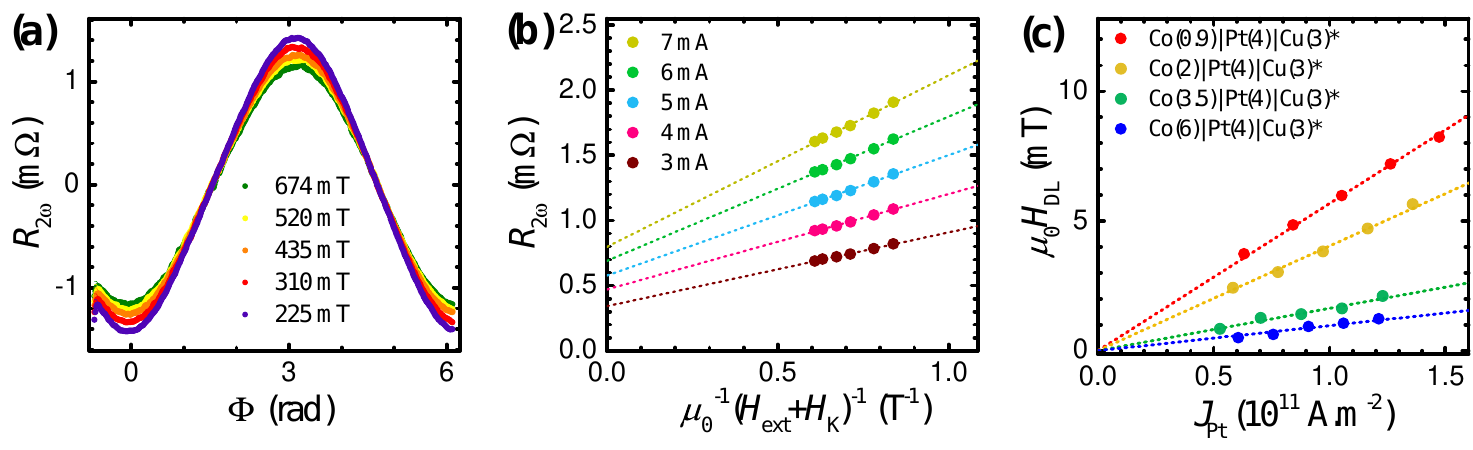}
\caption{\label{fig:Second} a) Azimuthal angle $phi$-dependence of the second harmonic transverse resistances $R_{2\omega}$ acquired on Co(2)|Pt(4)|Cu(3)* at different fields $H_{ext}$. b) Extraction of the $\cos\phi$ components of the second harmonic resistances $R_{2\omega}$ at different fields. c) Extraction of the damping-like (DL) effective field torque per density of current in Pt calculated with parallel resistance model. Measurements were performed at 300~K.}
\end{figure*}

The DL and FL components present different dependencies on the azimuthal angle allowing to be fitted separately. The DL fields may be disentangled from the thermal components ($I\alpha \Delta T$) from their strong external field dependence unlike the thermal component. In order to estimate the latter contribution, we have thus performed the measurements for different magnetic fields for Co(2)|Pt(4)|Cu(3)* as shown on Fig.~\ref{fig:Second}a for a current of 5~mA. We have extracted the amplitude of the respective $\cos\phi$ and $\left(2\cos^3(\phi)-\cos(\phi )\right)$ components corresponding to the respective DL and FL torques. In Fig.~\ref{fig:Second}b, we present the $\cos(\phi)$ DL components for which the thermal contribution corresponds to the y-intercept at the origin and the slope is equal to $R_{AHE}\times H_{DL}$. From R$_{AHE}$, we extract the effective DL as a function of the current density in Pt for the Co|Pt(4)|Cu(3)* series with Co thicknesses ranging from 0.9~nm to 6~nm (see Fig.~\ref{fig:Second}c) showing the expected linear relationship. Note that we systematically calculate the current density in Pt in order to compare the different samples integrating or not Cu* by using a parallel conductance model and from the analysis presented in the previous section. This method is equivalent to compare each measurement at constant applied electric field. The DL effective fields are fitted with a linear regression, the slopes of the fittings being equal to the ratio H$_{DL}$/J$_{Pt}$ and decrease with Co thickness in agreement with the expected evolution of the DL effective field ($\propto 1/t_{Co}$). 

\begin{figure*}
\includegraphics[width=0.75\textwidth]{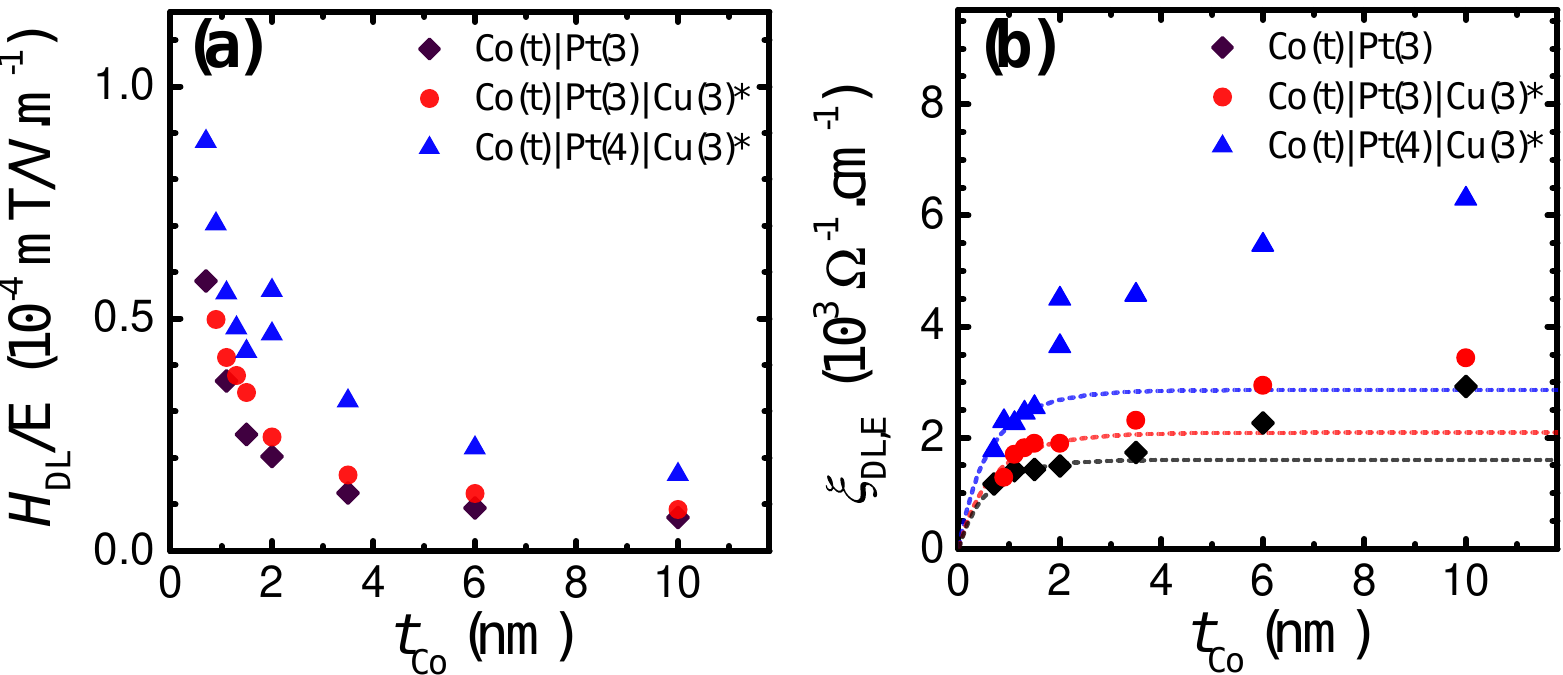}
\caption{\label{fig:DL}Co thickness dependence for the Co($t_{Co}$)|Pt(3), Co($t_{Co}$)|Pt(3)|Cu(3)* and Co($t_{Co}$)|Pt(4)|Cu(3)* series of a) damping-like (DL) torque effective fields over the electric field ratios, b) DL torque efficiency in unit of electric field. Measurements are performed at 300~K.}
\end{figure*}

\begin{figure*}
\includegraphics[width=0.75\textwidth]{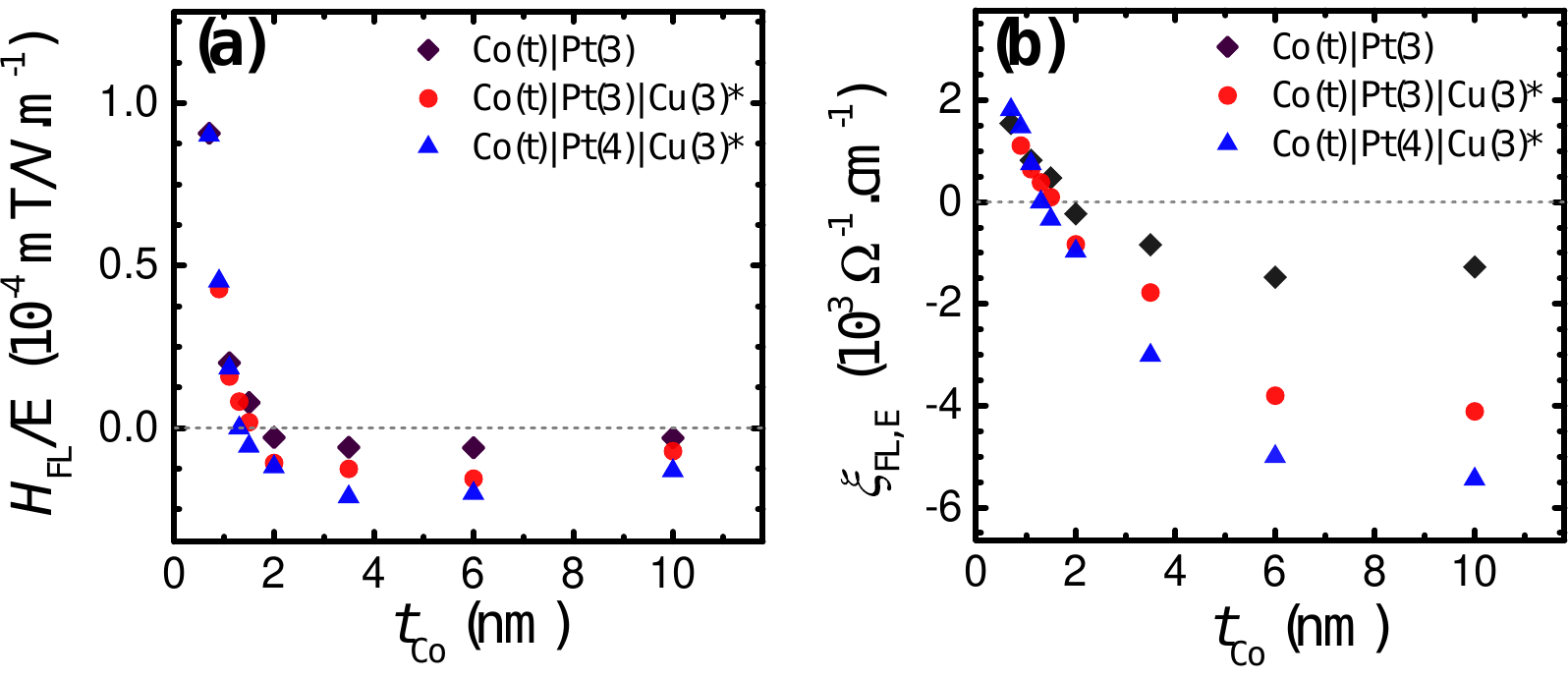}
\caption{\label{fig:FL}Co thickness dependence for the Co($t_{Co}$)|Pt(3), Co($t_{Co}$)|Pt(3)|Cu(3)* and Co($t_{Co}$)|Pt(4)|Cu(3)* series of a) Field-like torque (FL) effective fields on electric field ratios b) FL torque efficiency in unit of electric field. Measurements were performed at 300~K.}
\end{figure*}

\vspace{0.1in}

Following a similar procedure, the FL fields are extracted from the $2\cos^3(\phi)-\cos(\phi )$ component of the raw data obtained for different external magnetic fields. They are plotted \textit{vs.} 1/H$_{ext}$ and fitted with a linear regression as well where the slope is now equal to $R_{PHE}\times (H_{FL} + H_{Oe})$. However, the Oersted field component cannot be directly disentangled from the FL as they are both proportional to the current applied and both independent on the external field. They may be disentangled with further studies using thickness dependent measurements,  or comparison with models. Finally using the same estimation for the current densities in Pt we extract for each samples the ratios $(H_{FL}+H_{Oe})/J_{Pt}$. 

\vspace{0.1 in}

We carried out the Co thickness dependence of the SOT on \textit{i}) Co(t$_{Co}$)|Pt(3) series as reference of a dominant spin contribution to the SOT, typically displaying a short decoherence length of the order of 1.2~nm~\cite{krishnia2023nano}, on \textit{ii}) Co(t$_{Co}$)|Pt(3)|Cu(3)* series providing a small enhancement compared to the reference, and on \textit{iii}) Co(t$_{Co}$)|Pt(4)|Cu(3)* series displaying the maximum of orbit-to-spin conversion with a two fold enhancement. In the two latter series, the OHE for Cu is expected to be negligible~\cite{salemi2022,kelly2024}, in particular for very thin metallic Cu layer as considered here. The incoming spin current penetrating the thin Co ferromagnet may be generally composed of two terms of distinct origin: one from the SHE produced inside Pt and one from the orbital current. Such orbital current can be generated either in Pt, either by the OREE in Cu* and afterwards transmitted through Pt to reach Co. We have thus possibly four different channels to consider: \textit{i}) A spin-Hall current from  Pt, \textit{ii}) a secondary spin-current in Pt generated from orbital-to-spin conversion, \textit{iii}) a pure orbital current from Pt and  \textit{iv}) the orbital current from Cu* transmitted up to Co through the entire structure. The two main contributions can be gathered into the phenomenological equation: 

\begin{eqnarray}\label{eq_theta_orbit_to_spin}
\theta^{eff}=\theta_{Pt}^{SHE}+\eta_{L\rightarrow S}^{Pt}~\theta_{Cu}^{OREE}
\end{eqnarray}
with equivalent expression for the effective Hall conductivities ($\eta_{L\rightarrow S}^{Pt}$ is the characteristic efficient orbit-to-spin conversion coefficient). This expression has been recently considered to explain the enhancement of torque amplitudes and spin pumping signals integrating orbital-to-spin conversion processes~\cite{Ding2020, krishnia2024quantifying, Santos2023}. However it has to be noticed that only spin effects are considered in the net effective torque. As a matter of fact, to be more general, the possible injection of pure orbital current directly in the ferromagnet, that would not have been converted into spin current in the Pt spacer, have to be taken into account. We already know that processes \textit{i}) and \textit{ii}) previously described, exist (see \textit{e.g.} Ref.~\cite{krishnia2024quantifying}). A detailed examination of the decoherence length for the torque may help to confirm or not if processes \textit{iii}) and \textit{iv}) may subsist. 

\subsection{\label{sec:level4}Cobalt thickness dependence of DL torques}

We first discuss the evolution of DL torque with Co thickness in the three series. In Fig.~\ref{fig:DL}a, We compare reference sample Co($t_{Co}$)|Pt(3) (black diamond) to Co($t_{Co}$)|Pt(3)|Cu(3)* (red circles) and Co($t_{Co}$)|Pt(4)|Cu(3)* (blue triangles) by extracting the corresponding amplitude of the $\cos(\phi)$ DL component. This includes the \textit{i}) removal of the thermal component and \textit{ii}) the measurement of the $H_{DL}$ at different electric field $E$. The advantage of this method compared to a normalization \textit{vs.} the current density in Pt ($J_s^{Pt}$) as performed in Ref.~\cite{krishnia2024quantifying} is that, this way, the effect of the variation of the conductivity of each layer \textit{vs.} the film thickness are correctly considered. In particular, one can observe that the Co($t_{Co}$)|Pt(4)|Cu(3)* series (blue triangles) display a clear torque enhancement visible for $t_{Co} > 1$~nm for the Pt(4) sample.  We can compare the measured values to our previous results obtained on a different series of equivalent samples~\cite{krishnia2024quantifying}, where $t_{Co}$ was fixed to $t_{Co} = 2$~nm. Indeed, the $H_{DL}/J_{Pt}$ ratios at 2~nm Co in this work can then be estimated \textit{vs.} current density in Pt to give $(H_{DL}/J_{Pt})_{Co2|Pt3} = 1.56 $~mT/(10$^{11}$ A~m$^{-2}$), $(H_{DL}/J_{Pt})_{Co2|Pt3|Cu3^*} = 1.67$~mT/(10$^{11}$ A~m$^{-2}$) and $(H_{DL}/J_{Pt})_{Co2|Pt4|Cu3^*} = 3.92$~mT/(10$^{11}$ A~m$^{-2}$) in agreement with our previous values given in Ref.~\cite{krishnia2024quantifying}. Moreover, the DL torque normally decreases as Co becomes thicker as expected from its well-known $\tau \propto 1/t_{Co}$ dependence for sufficiently thick absorbing layer (the torque is purely interfacial at this scale).

\vspace{0.1in}

It is even more appropriate to consider the so-called effective spin-Hall conductivity for the torque defined as:
\begin{eqnarray}
\label{eq:effi}
  \xi_\text{DL,FL} = \frac{2eM_{s}t_{Co}H_{DL,FL}}{\hbar E}
\end{eqnarray}
with $H_{DL}$ the extracted DL field ($H_{FL}$ is the FL field), $M_{s}$ the saturation magnetization measured by SQUID magnetometry. The resulting $\xi_{DL}$ \textit{vs.} $t_{Co}$ are reported in Fig.~\ref{fig:DL}b. First, for $t_{Co}<2$~nm corresponding to a pretty short decoherence length for the three sample series, we deduce a dominant spin contribution, as expected. In this thickness range, the experimental data can be fitted with conventional 'spin-only model' (next section) corresponding to the absorption of the transverse spin-current $J_s^{\perp,in}$ by Co and showing a saturating behavior at thick Co. The value obtained for $\xi_{DL}$ for Co|Pt bilayers of about 2550~$\Omega^{-1}$/cm is in the range of what has been found previously in spin-pumping experiments when one considers spin-backflow effects and spin-memory loss. The difference in the saturation values of fitting functions with the larger effective torque obtained for Co($t_{Co}$)|Pt(3-4)|Cu(3)*, witnesses a second additive spin channel played by the orbital-spin conversion in Pt.

\vspace{0.1in}

We now discuss the data in the region corresponding to thicker $t_{co}$. We clearly observe a linear increase of the torque efficiency $\xi_{DL}$ for the three samples up to $t_{Co}=10$~nm. Such non-regular behavior, also highlighted in Ref.~\cite{ding2024mitigation}, has been assigned to the pure orbital contribution to the torque and arising in Pt~\cite{tanaka2008,kelly2024}. Our results confirm that the pure orbital torque may have a significant impact on the overall torque over a 10~nm thick Co, thus manifesting a pretty long orbital decoherence length in 3\textit{d} ferromagnet.

\subsection{\label{sec:level5}Co thickness dependence of Field-like torques}

In this section, we discuss the behavior of the FL torque. By using an equivalent procedure, we have extracted the $\left[2\cos(\phi)^{3}-\cos(\phi)\right]$ signal component from $V_{2\omega}$ signature of the total out-of equilibrium effective field $H_{FL}+H_{Oe}$, sum of FL torque and Oe fields. The FL plus Oe torques follow the same trends as the DL torque with an increase at small Co thickness (Fig.~\ref{fig:FL}a). Interestingly, the torque ratio $\frac{H_{FL+0e}}{H_{DL}}$ becomes larger than unity for $t_{Co}<1$~nm which manifests the smaller impact of the precession action in this thickness range smaller than the decoherence length~\cite{krishnia2023nano}.
Nonetheless, the Oe and FL fields can be hardly disentangled at this stage except from their different signatures upon the Co thickness. Indeed, whereas the Oe field is mainly independent on $t_{Co}$ at the nanometer scale, the FL torque is mainly interfacial. We report on the evolution of the $\left(H_{FL}+H_{OE}\right)/E$ ratio in Fig.~\ref{fig:FL}a for the three series. Additionally the FL torque present a sign change at $t_{Co}\approx 1.5$~nm indicating that $H_{Oe}$ has an opposite sign to the FL torque for 5\textit{d} Pt as previously demonstrated for 4\textit{d} Pd~\cite{gosh2017}. Note that a minimum is reached for each series at $t_{Co}=3.5$~nm, and the effective field then decreases in amplitude. This corresponds to the shunt of the top structure by the thick Co layer.

\vspace{0.1in}

The sum of FL and Oe spin-Hall effective conductivity extracted from Eqn.~[\ref{eq:effi}] are reported on Fig.~\ref{fig:FL}b for the three different series. We emphasize that the three series present similar values at low thicknesses $t_{Co}<1.5$~nm. The FL torque mainly manifests itself for Rashba-like contributions \textit{via} local exchange effects whilst less from spin current injection as investigated here. If we consider $H_{Oe}$ negligible for thin Co, FL torques are observed to be quite insensitive to any orbital contributions. In the region $t_{Co}>1.5$~nm, the three different series differs in agreement with a dominating Oe field torque contribution. Indeed, $H_{Oe}$ is expected to rise with the current asymmetry distribution, mainly flowing in the top layers of the stacks. Finally at large thickness ($t_{Co}>6$~nm) the efficiencies decrease because a dominant current in Co with a resulting more symmetric current path distribution.

\section{\label{sec:level6}Modeling of the transverse spin current penetration and DL torque}

In this section, we detail our quantitative model describing the spin-orbit torque arising from spin only in the limit of small $t_{Co}$, region wherein the orbital torque is negligible. Our approach to evaluate the out-of-equilibrium transverse spin accumulation $\Delta \mu_s (z)^\perp$ and spin current $J_s^\perp(z)=\sigma_F \frac{\partial \Delta \mu_s^\perp}{\partial z}$ ($z$ is the coordinate in the FM thickness layer) and transverse spin decoherence length $\lambda_\perp$ relies on the generalized complex Boltzmann equation describing the precession/relaxation intertwined mechanisms and parametrized by a complex spin resistance in Co~\cite{krishnia2023nano}. The general equation for $\Delta \mu_s (z)$ and $J_s^\perp(z)$ is then~\cite{petitjean2012,krishnia2023nano,manchon2024}:

\begin{eqnarray}
\left(\partial_z^2-\frac{1}{\lambda_{sf}^2}\right) \alpha_s^\perp-\frac{1}{\lambda_J^2}\alpha_s^\perp\times \mathbf{m}-\frac{1}{\lambda_\phi^2} \mathbf{m}\times\left(\alpha_s^\perp\times \mathbf{m}\right)=0
\end{eqnarray}
with $\alpha_S^\perp=\mu_s^\perp$ or $J_s^\perp$, $\lambda_{sf}$ the spin-diffusion length, $\lambda_J$ the Larmor precession length and $\lambda_\phi$ a certain memory length. We end up with a resolution of the form $\mu_s^\perp,J_s^\perp \propto A^{(+)} \exp{\left(+\frac{z}{\lambda_{F}^{*}}\right)}+A^{(-)} \exp{\left(-\frac{z}{\lambda_{F}^{*}}\right)}$ with $\lambda_{F}^{*}$ the complex decoherence length of the system; the real part corresponding to a vanishing propagation in the FM, whereas the imaginary part describes the effective precession length according to $\lambda_{F}^{*} = \lambda_{F}^{\bot,Re} + i\lambda_{F}^{L,Im}$.

\vspace{0.1in}

To solve the problem, we use the respective transmission and reflection coefficients at the boundaries with the SHE Pt layer in the the spin-mixing conductance ($G_{\uparrow \downarrow}$) framework at the Co|Pt interface. We find that the thickness dependence of the integrated torque in the FM is expressed as a function of the transverse spin-current really injected in Co. It results \textit{in fine} in:

\begin{eqnarray}
\sigma_{SHE}^{eff}=\sigma_{SHE}^{(0)}\frac{G_{\uparrow\downarrow}r_{s}^{Pt}\tanh\left(\frac{t_{Pt}}{2\lambda_{Pt}}\right)\tanh\left(\frac{t_{F}}{\lambda_{F}^{*}}\right)}{1+G_{\uparrow\downarrow}r_{s}^{Pt}\coth\left(\frac{t_{Pt}}{\lambda_{Pt}^{sf}}\right)\tanh\left(\frac{t_F}{\lambda^{*}_{_F}}\right)}
\end{eqnarray}
with $\sigma_{SHE}^{eff}$ the effective spin-Hall conductivity that we are searching for \textit{vs.} $t_{Co}$. The latter expression traduces the finite transmission coefficient (through $G_{\uparrow\downarrow}$), the spin-backflow process (through the inverse of the spin resistance $r_s^{Pt}$) and the finite thickness of both the heavy metal and the ferromagnet. In that sense, this formula typically describes the ferromagnet thickness-dependence of the spin-mixing conductance when $t_{Co}$ becomes smaller than the decoherence length. It has been established by considering perfect electronic reflection at the back-end of the thin ferromagnet. A possible impact of the spin memory loss at this interface may enter in an overall reduction prefactor. The real and imaginary parts correspond to \textit{integrated} DL and FL torques respectively over the ferromagnetic volume: 

\begin{eqnarray}
\tau_{DL}=\left(\frac{S\hbar}{m^*}\right)~ \text{Re}\left(J_{s}^{\bot,in}\right), \hspace{5mm}\tau_{FL}=\left(\frac{S\hbar}{m^*}\right) \text{Im}\left(J_{s}^{\bot,in}\right)
\end{eqnarray}
with $S$ the area of the Co|Pt interface. The fitting procedure has been performed by considering the product $G_{\uparrow\downarrow}r_{s}^{Pt} \approx 2 $ as well as the spin diffusion length in Pt $\lambda^{sf}_{Pt}=1.75 \pm 0.05$~nm~\cite{krishnia2024quantifying,krishnia2023nano}. It results in an effective spin-Hall conductivity of  $\sigma_{SHE}^{eff}\approx 1600~\left(\frac{\hbar}{e}\right)~\left(\Omega ~cm\right)^{-1}$ for pure Pt, $\sigma_{SHE}^{eff}\approx 2090~\left(\frac{\hbar}{e}\right)~\left(\Omega ~cm\right)^{-1}$ for Co($t_{Co}$)|Pt(3)|Cu(3)* and $\sigma_{SHE}^{eff}\approx 2860~\left(\frac{\hbar}{e}\right)~\left(\Omega~ cm\right)^{-1}$ for Co($t_{Co}$)|Pt(4)|Cu(3)*. These values are all in agreement with the ones previously reported on the same systems~\cite{krishnia2024quantifying} and extrapolated for slightly thinner Pt thickness (3~nm). The decoherence length extracted from $\xi_{DL}$ fittings show a behavior expected from SHE that is $\lambda^{\bot,Re}_{F} = 1.40\pm 0.16$~nm for the Co($t_{Co}$)|Pt(3) reference series. We find a short decoherence length as well for Co($t_{Co}$)|Pt(3)|Cu(3)* series with $\lambda^{\bot,Re}_{F} = 1.85\pm 0.20$~nm. For the Co|Pt(4)|Cu(3)* the extracted decoherence length is observed to reach $\lambda^{\bot,Re}_{F} = 1.67 \pm 0.10$~nm. The first result of the present study is the demonstration that the enhancement of the torque efficiency for $t_{Co}<2$~nm for Co(2)|Pt(4)|Cu* originates from spin effects due to a second spin channel due to the orbital-to-spin conversion in Pt as suggested above. The major result of this paper is the linear increase of $\xi_{DL}$ for $t_{Co}\geq 2$~nm assigned to pure orbital injection and interactions in Co.

\begin{figure}
\includegraphics[width=0.42\textwidth]{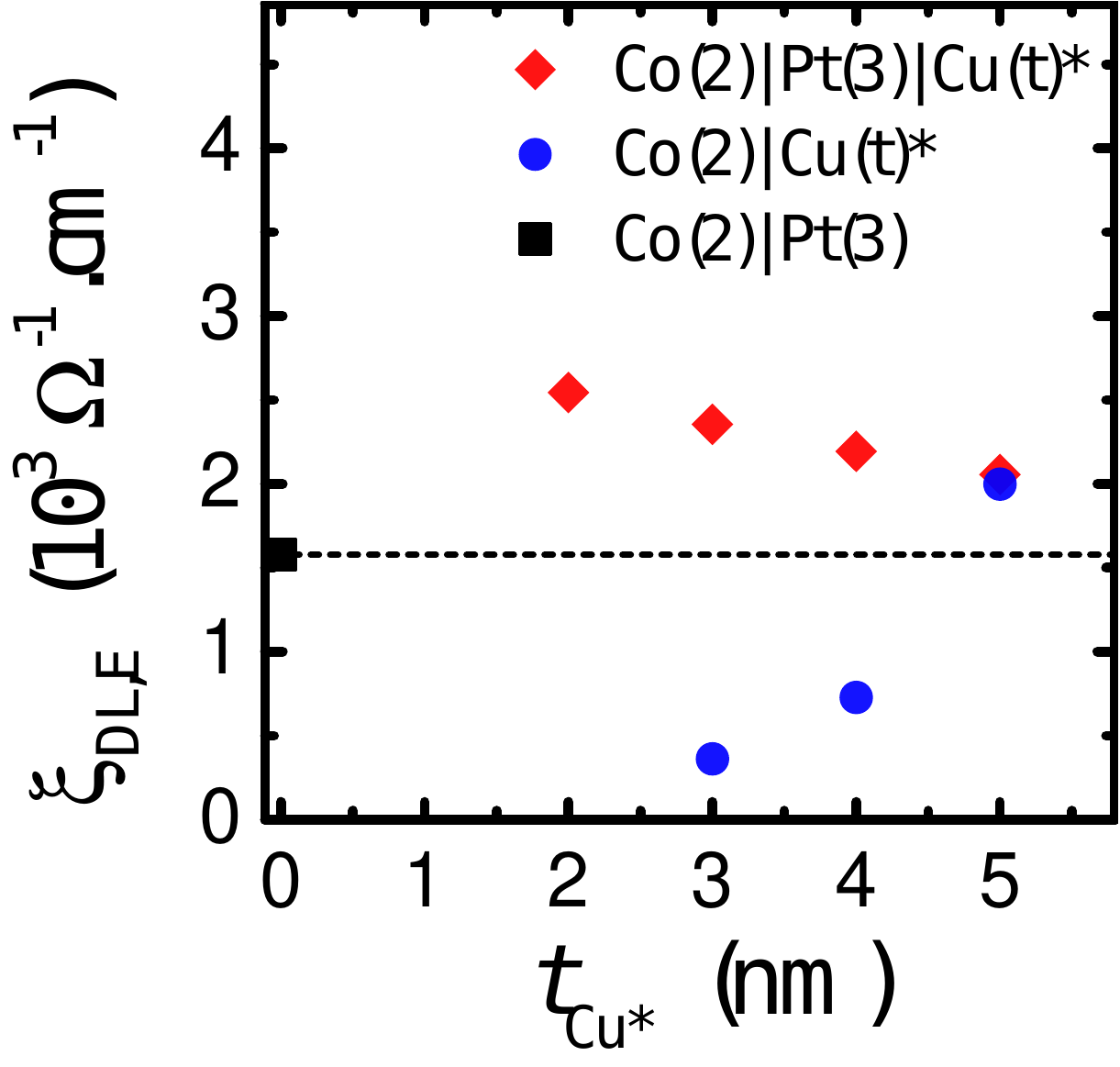}
\caption{\label{fig:Cu}Cu* thickness dependence DL effective field normalized by the electric field for light element system Co(2)|Cu(t$_{Cu}$)* and orbit-to-spin conversion system Co(2)|Pt(3)|Cu(t$_{Cu}$)*. Measurements were performed at 300~K.}
\end{figure}

\section{Orbital currents generation in naturally oxidized Cu (cu*)}

In this section, we investigate the Cu* thickness dependence on either Co(2)|Cu$(t_{Cu}$)*, a pure orbital torque system, and Co(2)|Pt(3)|Cu($t_{Cu}$)*, a system involving orbit-to-spin conversion in which moreover Pt plays the role of a diffusion barrier for oxygen~\cite{krishnia2023nano}.
In Fig.~\ref{fig:Cu}, we present the DL field extracted from our second harmonic analysis (red losange). We have normalized the data by the electric field applied $H_{DL}/E$ for a fair comparison between samples. As expected, the Co|Pt|Cu* series has a maximum torque efficiency for the thinner Cu $t_{Cu*}=2$~nm reaching $\xi_{DL}=2.5\times 10^3 ~\left(\frac{\hbar}{e}\right)~\left(\Omega~ cm\right)^{-1}$, and hereafter gradual decreasing values down gradually in the Cu thickness window up to $t_{Cu^*}=5$~nm ($\xi_{DL}=2.1\times 10^3~ \left(\frac{\hbar}{e}\right)~\left(\Omega~ cm\right)^{-1}$). It has to be emphasized that this values obtained \textit{in fine} for $t_{Cu}=5$~nm lies slightly over the reference value obtained for Co|Pt previously found (small black square). Such behavior also discards any bulk OHE from metallic Cu to confirm an interfacial OREE. Indeed, in tat case, the current density in Cu*, and accordingly the torque efficiency $\xi_{DL}$, would be expected to largely increase with $t_{Cu*}$ contrary to what we find.

\vspace{0.1in}

Finally, we analyze the results for the Co|Cu* series in which we find a torque increase \textit{vs} the Cu thickness in the $t_{Cu*}=3-5$~nm range (blue circle), starting from relatively low values, $\xi_{DL}=0.36\times 10^3~\left(\frac{\hbar}{e}\right)~\left(\Omega~ cm\right)^{-1}$, reaching the same value of  $\xi_{DL}=2.0\times 10^3 \left(\frac{\hbar}{e}\right)~\left(\Omega~ cm\right)^{-1}$  as that found for Co(2)|Pt(3)|Cu(5). Importantly, we note that the orbital torque from pure light element, presently, Co(2)|Cu(5)* exceeds the one of Co(2)|Pt(3) confirming thus the strong impact of the orbital currents on the torque. Finally, for these series the FL torque are found negligible as expected for $t_{Co}=2$~nm . 

\section{\label{sec:Disc}Conclusions}

In conclusion, we have demonstrated an enhancement of the damping-like (DL) torque in Co|Cu* and Co|Pt|Cu* sample series with the benefit of additional angular momentum channels to the SOT. Those are played either by additional pure orbital current (OT) generated at the Cu|CuO$_x$ interface (Cu*) or by orbital-to-spin conversion of those orbital currents into spin in interlayer Pt.
In details, our results and analyses indicate that the observed torque enhancement (maximum of twofold enhancement) in the region of thick Co ($t_{Co}\geq 2$~nm) is due to the action of a pure orbital current contrary to what is expected by the orbit-to-spin conversion in Pt relevant at smaller Co thickness ($\simeq 1-2$~nm). Such orbital current is generated by OREE at the Cu|CuO$_x$ interface and partially transmitted through Pt until Co.
The second issue deals with the possibility to consider other mechanisms at play in addition to OREE at Cu|CuO interfaces discarded in the present analysis. To go further in the understanding, using other 3\textit{d} ferromagnets with larger spin-orbit coupling, as Ni or CoPt, could be investigated in the future. In particular, Ni presents a higher sensitivity to the orbital current injection and thus would allow a more accurate analysis. Finally reproducing FM thickness dependence with other measurements (STT-FMR or current-induced magnetization switching) could bring additional information. Finally we also demonstrate the interfacial nature of the orbital current generation in the top naturally oxidized Cu layer, through the Cu thickness dependence of Co(2)|Pt(3)|Cu(t$_{Cu}$)* series presenting constant values, contrary to the Co(2)|Cu(t$_{Cu}$)* series that exhibit an increase with Cu thickness that we may explain with crystalline and chemical arguments. 

\begin{acknowledgments}
This research was supported by the EIC Pathfinder OPEN grant 101129641 “OBELIX”, by grant ANR-20-CE30-0022-01 “ORION” and by ANR grant "STORM" (ANR-22-CE42-0013-02). We thank M. Pannetier-Lecoeur and C. Fermon (CEA-SPEC, Gif/Yvette) for fruitful discussions.
\end{acknowledgments}

\vspace{0.1in}

\section*{Data Availability Statement}

The data that support the findings of this study are available from the corresponding author upon reasonable request.

\section*{References}

\nocite{*}
\bibliography{biblio2}

\end{document}